\documentstyle[twocolumn,twoside,prl,floats,aps,epsfig]{revtex}

\newcommand{\eq}{\begin{equation}}
\newcommand{\en}{\end{equation}}
\newcommand{\eqa}{\begin{eqnarray}}
\newcommand{\ena}{\end{eqnarray}}

\newcommand{\ZZ}{\hbox{{\rm Z{\hbox to 3pt{\hss\rm Z}}}}}

\def\bea{\begin{eqnarray}}
\def\eea{\end{eqnarray}}
\def\be{\begin{equation}}
\def\ee{\end{equation}}

\def\d{\partial}
\def\D{\delta}

\def\A{{\cal A}}
\def\s{\sigma}
\def\l{\lambda}

\def\t{\Delta}

\def\mycaptionl#1{%
\refstepcounter{figure}
\begin{center}
\hskip 1pt\vskip -0.6cm
\begin{minipage}{12cm}
\small {\bf Fig. \hskip -3pt\arabic{figure}}: {\sl #1}
\end{minipage}
\null\hskip 1pt\vskip -0.2cm
\end{center}}

\tighten
\begin{document}
\draft
\vskip 0.4cm

\title{  
 Bound states in the three-dimensional $\phi^4$  model.}

\author{
M. Caselle$^a$
M. Hasenbusch$^b$ P. Provero$^{c,a}$ and K. Zarembo$^{d,e}$}

\address{$^a$  Dip. di Fisica 
Teorica dell'Universit\`a di Torino and I.N.F.N.
 via P.Giuria 1, I-10125 Turin,Italy\\
$^b$ Humboldt Universit\"at zu Berlin, Institut f\"ur Physik
 Invalidenstr. 110, D-10099 Berlin, Germany\\
 $^c$ Dip. di Scienze e Tecnologie Avanzate,
Universit\`a del Piemonte Orientale, I-15100 Alessandria, Italy\\
 $^d$ Dep. of Physics and Astronomy, University of British
Columbia, Vancouver, BC V6T 1Z1, Canada\\
$^e$ Institute of Theoretical and Experimental Physics, 
B. Cheremushkinskaya 25, 117259 Moscow, Russia\\
\parbox{14 cm}{\medskip\rm\indent
We discuss the spectrum of the three dimensional 
$\phi^4$ theory in the broken symmetry phase. In this phase the effective
potential between  the elementary quanta of the model is attractive and 
bound states of two or more of them may exist. We give theoretical and
numerical evidence for the existence of these bound states. Looking in 
particular at the Ising model realization of the $\phi^4$ theory we show,
by using duality, that
these bound states are in one-to-one correspondence with the glueball states of
the gauge Ising model. We discuss some interesting consequences of this
identification.
\\
Preprint DFTT 4/2000, HU-EP-00/09, ITEP-TH-84/99.
}}
\maketitle
\narrowtext
Three dimensional
statistical systems with global $Z_2$ symmetry, the Ising model
being the classic example, lie in the 
universality class of the $\phi^4$ field theory. Critical phenomena
in such systems are known to be accurately described by simple
perturbative methods  \cite{zin96}.
Given the success of perturbative methods, the appearance of excited states
in the broken symmetry phase of the critical Ising model and in the
3D $\phi^4$ theory, which were
found in \cite{chp99,chp99b},
comes as a surprise, since 
scalar field theory apparently describes only one particle
as long as  interactions can be treated perturbatively. We shall argue that 
this is not the case and there is  room for a rich spectrum 
of excitations in the broken symmetry phase of the $\phi^4$
theory even if the interaction is weak.
\par
The excited states show up as poles of the 
correlation functions in the complex momentum plane and give visible 
contribution to certain universal quantities. 
The first excited state lies just below the two-particle threshold: its 
mass is $M=1.83(3)m$ \cite{chp99},
where $m$ is the mass gap. 
The closeness of $M$ to the threshold suggests the interpretation of 
this excitation as a weakly coupled bound state of two elementary excitations. 
\par
Indeed, 
the two-particle forces are attractive in 
the broken-symmetry phase of the $\phi^4$ 
theory, and  
bound states of two or more
elementary quanta  may in principle be formed. 
In four dimensions,
these states indeed exist in the low-temperature
regime, but disappear as the continuum limit is 
approached~\cite{lw88,gl98},
in agreement with triviality.
\par
In this letter we address the three-dimensional case.
Numerical simulations show
 that non-perturbative states survive the continuum limit
in 3d \cite{chp99,chp99b}.
We shall argue that these states can be identified
with the multiparticle bound states. By considering the Ising 
realization of the $\phi^4$ model and using duality
we shall also show that there is an exact one-to-one mapping between the
 bound states of the Ising model (and hence, thanks to universality,
also of the $\phi^4$ theory) and the glueball states of the gauge 
Ising model. 

\begin{figure}[h]
\hspace*{1.5cm}
\epsfig{file=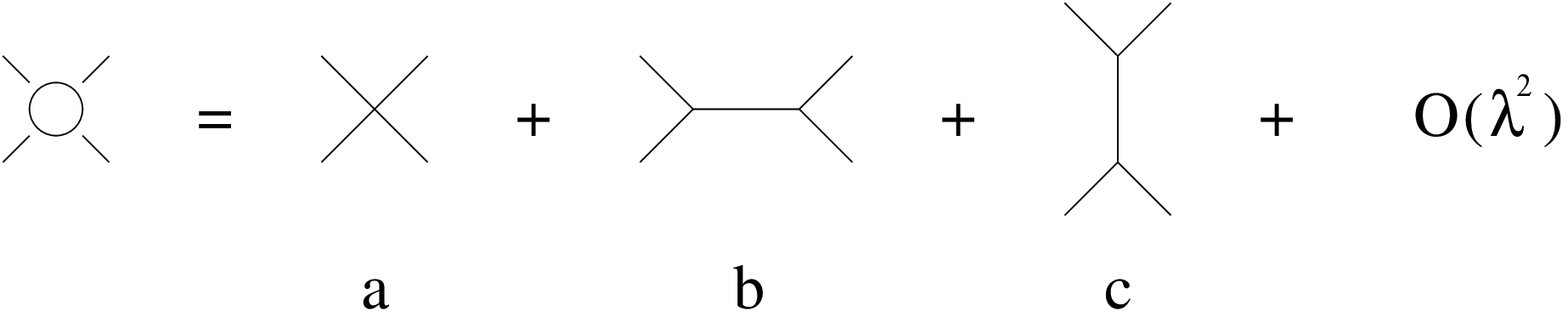,height=6cm,angle=-90}
\caption[x]{\small Feynman diagrams contributing to the $2\rightarrow 2$
amplitude at the leading order in $\l$.}
\label{kernel}
\end{figure}

\vskip0.5cm
{\bf Bound states in the $\phi^4$ theory.}
We consider the $\phi^4$ theory: 
$$S=\int d^3x\,\left[\frac12\,(\d\phi)^2+\lambda(\phi^2-v^2)^2\right].$$
The field $\s=\phi-v$ acquires 
the mass $m^2=8\lambda v^2$ at the tree level and
is reasonably weakly coupled in the critical regime \cite{zin96}, since
the critical value of the dimensionless interaction constant, 
$\lambda/m$, is not too big. The forces between
elementary quanta of
the field $\s$ are attractive: This can be shown 
by inspecting the scattering of two 
non-relativistic particles. 
There are three leading-order diagrams contributing to 
this process, shown in Fig.~\ref{kernel}. 
The contact interaction, diagram $(a)$, 
contributes $-12\l$ to the scattering amplitude. 
As a first approximation, we can neglect 
altogether the momentum flow in diagrams $(b)$ and $(c)$.
In this way, diagrams $(b)$ and $(c)$ contribute $-12\l$ and $72\l$, 
respectively.
Collecting the three terms together, we get for the amplitude
$\A=48\l$.
The positive sign of the amplitude means that the particles attract 
each other.
\par
In this limit, the non-relativistic Hamiltonian describing the interaction of
two particles is
\be
H=\frac{{\bf p}_1^2}{2m}+\frac{{\bf p}_2^2}{2m}-\frac{12\l}{m^2}\,
\D({\bf x}_1-{\bf x}_2),
\ee
which
reproduces the field-theory 
scattering amplitude in the Born approximation \cite{albeverio}. 
Note the factor of $1/(2m)^2$,
which accounts for the relativistic normalization of the wave
functions in field theory.
The quantum-mechanical system of two
dimensional particles interacting via a $\delta$-type potential
 develops short-distance divergences and requires
a regularization \cite{t79,h92}. In our case, the cutoff is
proportional to $m$, because at momenta of order $m$ 
the non-relativistic approximation becomes inadequate.
\par
The binding energy $\t m$, $\t\ll 1$, is determined by the Schr\"odinger
equation  for the relative motion wave function:
\be
\left(-\frac{1}{m}\,\d^2+\t m\right)\psi({\bf r})=\frac{12\l}{m^2}\,
\psi(0)\D({\bf r}),
\ee
which, after the
Fourier transform, gives the consistency condition:
\be
1=\frac{12\l}{m^2}\int\frac{d^2p}{(2\pi)^2}\,\frac{1}{p^2/m+\t m}
=\frac{3\l}{\pi m}\ln\frac{\Lambda^2}{\t m^2}.
\ee
Taking $\Lambda^2=\kappa m^2$, we get for the binding energy:
\be
\t=\kappa\exp\left(-\frac{\pi m}{3\l}\right).
\ee
The constant $\kappa$ cannot be determined in the approximation used 
above and requires the inclusion of loop corrections and of the 
momentum dependence of diagrams $b$ and $c$. 
These corrections can be systematically taken into account
in the approach based on the Bethe-Salpeter equation, which will be
reported elsewhere \cite{chpz00};
here we only quote the result: 
$\kappa=4/\sqrt{3}$.
\par
The above discussion tells us that only one bound state of two elementary
quanta may exist in the broken phase of the $\phi^4$ model in three dimensions.
This is in accord with numerical simulations 
\cite{chp99,chp99b}.
However, a rich spectrum of bound states, with different values of the angular
momentum, can be found if we look at
the bound states of three or more elementary quanta.
Bound states of $n\ge 2$ particles could be studied in principle 
within the Bethe-Salpeter approach, 
but even the non-relativistic approximation described above
becomes too complicated
as the number of particles involved in the bound state increases. 
Up to our knowledge, the only existing result 
in the literature is a discussion of the $0^+$ bound state of three particles 
which can be found in~\cite{bt79}. The counterpart of this state is
also seen in the numerical simulations \cite{chp99b}. 
An easier way to understand the qualitative features of the bound states
is to study the Ising model in the low-temperature phase far below
criticality.
\vskip0.5cm 
{\bf The Ising model at low temperature.}
To proceed in understanding the structure of the bound state spectrum in the
$\phi^4$ theory, let us address the same problem in the case of the 
low temperature phase of the 3d Ising model. 
 The Ising model and the $\phi^4$ theory belong to the same
universality class. 
Therefore they should have the same spectrum in the critical limit.
Indeed,  
in~\cite{chp99,chp99b} it was shown in a Monte Carlo study
that 
the two models share the same spectrum of non-perturbative
states.

The main advantage of working with the Ising model
is that the spectrum can be analyzed in a low temperature expansion of the 
transfer matrix (see~\cite{FC71,C73}). The starting point of this expansion
is to ignore the 
interactions between time-slices.
In this approximation, the 
vectors that correspond to a single configuration on a time-slice 
become eigenvectors of the transfer matrix. The eigenvalues of the transfer 
matrix
are directly 
given by the number of frustrated bonds.
\par 
In this framework, the bound state of
two particles is obtained by flipping two nearby spins.
If we flip two spins
which are separated by a distance of more than one lattice spacing,
the total number of frustrated bonds is exactly twice that of a single 
particle.
On the contrary, if we flip two nearby spins,  
the number of frustrated bonds is
reduced by two.
This difference is the binding energy of
the bound state. The fact that we are constrained to choose the two spins in
nearest neighbor sites is another way to state that the attractive force
between the two particles has a very short range. 
This procedure can be iterated, and one can construct clusters of $k$ nearby
flipped spins which have a non-zero binding energy and are related to  bound 
states of higher mass.
\par
It is also possible to select bound states of  non-zero
angular momentum. These combinations can
be constructed by using standard group-theoretical techniques.  They are
discussed in~\cite{acch96}. Let us only recall here two results which are of
interest for the present analysis.
\par
On a (2+1) dimensional lattice the group of rotations and parity reflections 
is reduced to the dihedral group $D^4$ which has four one-dimensional
and one two-dimensional irreducible representations. 
The $0^+$ state is associated to the trivial one-dimensional 
irreducible representation.
The $2^+$ and $2^-$ states are degenerate and correspond to two other 
one-dimensional representations. The simplest possible realization of
the $2^-$ state is represented in Fig. 2. At least three flipped spins are 
needed
to create such a state (the $2^+$ state could be also realized in a simpler 
way, but a general theorem forces its mass to be the same of the $2^-$ one in  
the continuum limit). Thus we expect that this state should appear as a bound
state of at least three elementary quanta.
The last one-dimensional irreducible representation corresponds to the 
$0^-$ state.
The simplest possible representation of this
state is reported in Fig. 3 and requires at least four flipped spins.
Finally, all the states with odd angular momentum 
are collected in the two dimensional representation. As in
the $2^+$,$2^-$ case they are all degenerate in parity. 
\par
Let us summarize the pattern of bound states as it emerges from these  
considerations. With two elementary quanta we may only
create a bound state with quantum numbers $0^+$. We shall denote it with 
$0^{+,*}$ to distinguish it from the single particle excitation which has the 
same quantum numbers. With three quanta we may create a pair of bound states 
$2^{+}$ and $2^{-}$ and a new $0^+$ excitation that we shall call $0^{+,**}$. 
With four particles we shall have a $2^{\pm,*}$ pair, a $0^{+,***}$ state and 
a new state with quantum numbers $0^-$. With five particles a new pair of 
states of the type $1^{\pm}$ appears, and so on.
\par
Let us now make the crucial assumption that the binding energy is always
much smaller that the mass of the constituent particles: then the mass 
of each bound state will be essentially given by the number of particles
needed for its formation, minus a small correction given by the binding
energy. In this way one obtains
a detailed prediction of the qualitative features of the spectrum, based 
only on the interpretation of the states as bound states and the 
group-theoretical facts described above. 
\par
Numerical simulations \cite{chp99,chp99b,chpz00,acch96} 
show that not only these predictions
are fulfilled, but that the same qualitative features of the spectrum 
survive well beyond the low-temperature regime and into the scaling region. 
Connected correlators of several composite operators
are computed in Monte Carlo simulations and used to extract the spectrum
in the various angular momentum channels. 
The measured masses
exactly follow the pattern suggested above.
This is a strong indication that
the spectrum is indeed made of bound states of the elementary quanta, and 
that these bound states survive in the continuum limit.
\par
The fact that states with angular momentum $2$ are lighter than that with
angular momentum $0^-$ is rather unexpected in standard quantum field theory.
However it is a well established feature of the
glueball spectrum in (lattice) gauge theories. This is the first 
hint that the spectrum of bound states of the 3d Ising model has 
something to do with the glueballs of gauge theory. The reason for 
this is obviously the duality between spin model and gauge model, 
that we will now discuss.
\vskip0.5cm
\begin{figure}
\begin{center}
\null\hskip 1pt
\epsfxsize 4cm
\epsffile{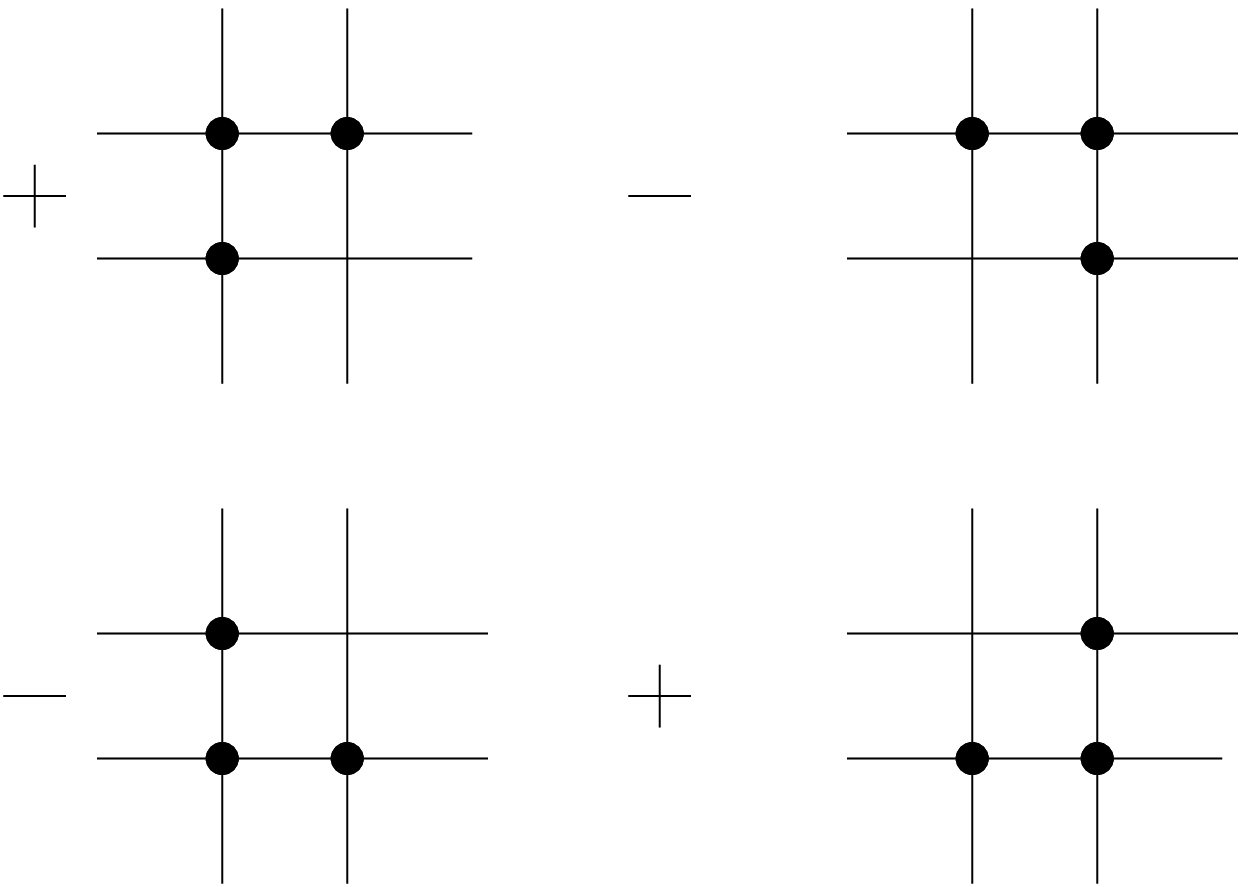}
\vskip 0.15cm\noindent
\end{center}
\mycaptionl{Operators for $2^-$.}
\end{figure}

\begin{figure}
\begin{center}
\null\hskip 1pt
\epsfxsize 8cm
\epsffile{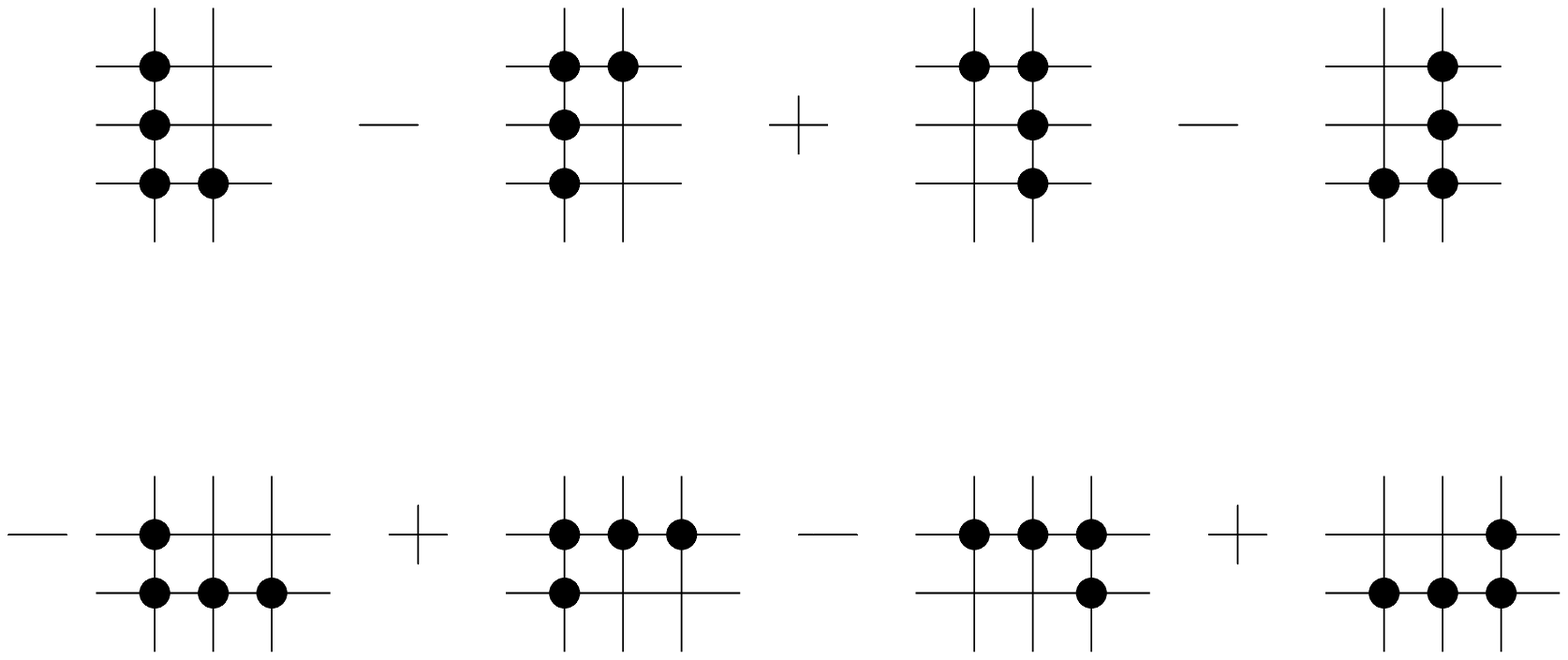}
\vskip 0.15cm\noindent
\end{center}
\mycaptionl{Operators for $0^-$.}
\end{figure}
{\bf Duality.}
Duality is usually expressed as an exact equality between partition functions 
in infinite volume, 
hence in
principle it does not automatically implies that the two theories must have the
same spectrum. However it can be shown that duality holds not only in the
thermodynamic limit {\sl but also for finite lattices}. This 
correspondence is
not trivial and requires a careful analysis of the boundary
conditions of the two models~\cite{chpz00}. 
Since the approach to the thermodynamic limit of the finite
volume partition function is driven by the full spectrum of excited states of
the theory, the finite volume duality implies that the spectra of the two 
models must coincide. In particular the bound state of quantum numbers $J^P$ of
the Ising spin model coincides (hence has exactly the same mass) with the
$J^P$ glueball of the gauge Ising model. This identification has two
interesting consequences. 
The first one is that the Bethe-Salpeter approach to the calculation of
bound states in $\phi^4$ theory, described above, becomes an analytical
tool to evaluate the masses of 
the first states of the glueball spectrum 
of the gauge Ising model. In principle
(apart from technical difficulties) this could be extended to the whole
glueball spectrum, and represents a powerful alternative to the Isgur-Paton
model, which in the case of the Ising gauge model gives rather poor 
result~\cite{acch96}.
A second, more important consequence of this identification is that it gives a
possible explanation for a peculiar degeneracy observed in the Monte Carlo
estimates of the glueball masses in the 3d gauge Ising model~\cite{acch96} for
which no alternative explanation exists.
This intriguing feature of the spectrum 
can be immediately appreciated by
looking at Tab.~1, (data taken from~\cite{acch96,t98}). 
In Tab.~1 the asterisks denote
the radial excitations, thus $0^+$ is the lowest state in the family with
quantum numbers $0^+$, $0^{+,*}$ the next one and  $0^{+,**}$ the third one. 
$0^+$ is related by duality to the single-particle state of the 3d Ising model,
$0^{+,*}$ to the first bound state and so on. 
The degeneracy involves the
pairs $(0^{+,**},2^{\pm})$, $(0^{-},2^{\pm,*})$, $(0^{-,*},1^{\pm})$ (the
last one is only roughly established, it holds within the errors).
Let us stress that
this degeneracy has no obvious physical reason. The only one which we would
expect on physical grounds is the one
 between $J^+$ and $J^{-}$ states (for $J\neq
0$) (see~\cite{acch96} for a discussion of this point)
 which is indeed present and has been already taken into account in Tab.~1.
Moreover it {\em is not} explained by the Isgur-Paton model (last column
of Tab. 1). This degeneracy seems to be  a rather
 deep phenomenon since it is also present in the glueball spectrum of the SU(2)
model in (2+1) dimensions.
In the third column of Tab. 1 we report the data on SU(2)
obtained by Teper~\cite{t98} (the underlined values are our extrapolations of
the finite-$\beta$ values reported in~\cite{t98}). One can easily see that
the same pattern of
degeneracy is present both in the SU(2) and in the Ising gauge spectra.
On the contrary, all these degeneracies seem to be
lost in  the $SU(N)$, $(N>2)$ case (see the data reported in~\cite{t98}).
\par
This degeneracy is well explained by the interpretation of the glueballs
as bound states of the dual spin model: the degenerate glueball states 
are simply bound states of the {\em same number} $n_c$ of constituent 
particles, namely $n_c=3,4,5$ respectively for the 
$(0^{+,**},2^{\pm})$, $(0^{-},2^{\pm,*})$ and  $(0^{-,*},1^{\pm})$ 
degeneracies.
In fact, according to the assumption stated above,
 the major contribution to the mass of the bound state is given by the
number of elementary quanta involved. The dependence on the various quantum
numbers is encoded in 
the binding energies $\Delta$ which however give only a small correction to 
the mass. This results in the approximate degeneracies observed in the
simulations. Notice that we do not expect to have {\sl exact} degeneracies,
since there is no reason to expect 
the binding energy 
to be
exactly the same for
different bound states.
\begin{table}[ht]
\label{ip5f}
\caption{\sl Comparison between the  Ising, SU(2) and Isgur-Paton spectra. The
masses are measured in units of the string tension.}
  \begin{center}
\begin{tabular}{|l|c|c|c|}
\hline
$J^P$ &  Ising & SU(2) & IP    \\
\hline
$0^+$  &  3.08(3)& 4.718(43) & 2.00  \\
$0^{+,*}$ & 5.80(4)& 6.83(10)   & 5.94  \\
$0^{+,**}$  &  7.97(11)& 8.15(15) & 8.35   \\
$2^{\pm}$  &  7.98(8) & 7.84(14)& 6.36  \\
$2^{\pm,*}$  &  9.95(20)&\underline{9.30(50)}  & 8.76  \\
$0^-$   &   10.0(5) & 9.95(32) & 13.82 \\
$0^{-,*}$  & 13.8(6) & \underline{11.30(80)}  & 15.05  \\
$(1)^\pm$   & 12.7(5)& 10.75(50) & 8.04  \\
\hline
\end{tabular}
  \end{center}
\end{table}
\vskip0.5cm
{\bf Conclusions.}
Our analysis shows that bound states are very likely to exist in the 
broken-symmetry phase of 3d $\phi^{4}$ and Ising models. Their 
existence can be inferred both from the Bethe-Salpeter equation of the 
field theory and the strong-coupling analysis of the spin model, and 
is strongly confirmed by numerical simulations. 
\par
Duality allows one to apply the same analysis to the glueball spectrum 
of the 3d Ising gauge model, which exactly coincides with the one of
the spin model. The interpretation of the latter as a spectrum of 
bound states provides a natural explanation for several  
features of the glueball spectrum, such as its peculiar dependence on 
the angular momentum and its characteristic degeneracies.
\vskip 1cm
{\bf  Acknowledgements.}
We would like to thank M. Campostrini, A. Pelissetto, P. Rossi and E. Vicari 
for fruitful discussions, and the organizers of the INTAS meeting 1999 
for providing the stimulating environment in which this work was started.
This work was partially supported by the 
European Commission TMR programme ERBFMRX-CT96-0045.
The work of K.Z. was supported by the PIMS Postdoctoral Fellowship and
by NSERC of Canada.
 
\end{document}